\documentclass{PoS}
\usepackage{cite}

\newcommand{\BBbar}{\ensuremath{B\bar{B}}}
\newcommand{\Bp}{\ensuremath{B^{+}}}
\newcommand{\Bz}{\ensuremath{B^{0}}}
\newcommand{\Bzb}{\ensuremath{\bar{B}^{0}}}
\newcommand{\YFS}{\ensuremath{\Upsilon(4S)}}
\newcommand{\Ep}{\ensuremath{e^{+}}}
\newcommand{\Em}{\ensuremath{e^{-}}}
\newcommand{\pip}{\ensuremath{\pi^{+}}}
\newcommand{\piz}{\ensuremath{\pi^{0}}}
\newcommand{\pim}{\ensuremath{\pi^{-}}}
\newcommand{\Mbc}{\ensuremath{M_{\rm bc}}}

\newcommand{\De}{\ensuremath{\Delta E}}
\newcommand{\Dt}{\ensuremath{\Delta t}}
\newcommand{\phitwo}{\ensuremath{\phi_{2}}}
\newcommand{\aone}{\ensuremath{a_{1}(1260)}}
\newcommand{\Acp}{\ensuremath{{A}_{CP}}}
\newcommand{\Scp}{\ensuremath{{S}_{CP}}}

\newcommand{\BtoDK}{\ensuremath{B^0 \to D^0[K_S^0\pip\pim]K^{*0}}}

\definecolor{mygrey2}{rgb}{0.4,0.4,0.4}

\definecolor{myorange}{rgb}{0.9,0.6,0}
\definecolor{mygrey2}{rgb}{0.4,0.4,0.4}
\definecolor{mygrey}{rgb}{0.25,0.25,0.25}
\definecolor{mygreen}{rgb}{0.,0.35,0.}

\title{Recent studies of CP violation in bottom meson decays at Belle}

\ShortTitle{Recent studies of CP violation in bottom meson decays at Belle}

\author{\speaker{P.~Vanhoefer}\thanks{for the Belle Collaboration}\\
        Max-Planck-Institut f\"ur Physik, 80805 M\"unchen\\
        E-mail: \email{pvanhoef@mpp.mpg.de}}

\abstract{We present a summary of recent studies on $CP$ violation with the Belle experiment using the final data sample of $772 \times 10^{6}$ \BBbar\ pairs produced at the \YFS\ resonance at the KEK asymmetric \Ep\Em\ collider. We discuss preliminary measurements of the branching fraction, the polarization and the $CP$ asymmetries of the decay $B^0 \rightarrow \rho^{+} \rho^{-}$ and an updated constraint on the CKM angle $\phi_2$ from the $B\to\rho\rho$ system. Being also related to $\phi_2$, we present a preliminary measurement of the branching fraction of $B^0\to\piz\piz$ decays. Last, a preliminary model independent dalitz plot analysis of the decay $B^0\to D^0[K_S^0\pip\pim]K^{*0}$ is presented and its sensitivity to the CKM angle $\phi_3$ is discussed.}

\FullConference{The European Physical Society Conference on High Energy Physics\\
		22--29 July 2015\\
		Vienna, Austria}

\begin{document}
\section{Introduction}
Testing the predictions of the Cabibbo-Kobayashi-Maskawa (CKM) mechanism for violation of the combined charge-parity ($CP$) symmetry~\cite{C,KM} is one of the major precision tests~\cite{JPsiKs_theo,ccK_Belle,BigiSander2} of the flavor sector of the Standard-Model (SM). The Belle experiment at KEK significantly contributed to the validation of the CKM scheme and to constraining the unitarity triangle for $B$ decays to its current precision. Up-to-date, the SM describes the data impressively well, but there is still room for a deviation from unitarity, which clearly would point towards physics beyond the SM. 
These proceedings give a summary of the experimental status of measurements related to two of the CKM angles defined from the CKM matrix elements as: $\phitwo \equiv \arg(V_{td}V^{*}_{tb})/(-V_{ud}V^{*}_{ub})$ and $\phi_3 \equiv \arg(V_{ud}V^*_{ub})/(-V_{cd}V^*_{cb})$. All measurements presented here are based on Belle's final data set of $772 \times 10^{6}$ $B\bar{B}$ pairs.

\section{The CKM Angle \phitwo}

\begin{figure}[htb]
  \centering
  \includegraphics[width=.5\columnwidth]{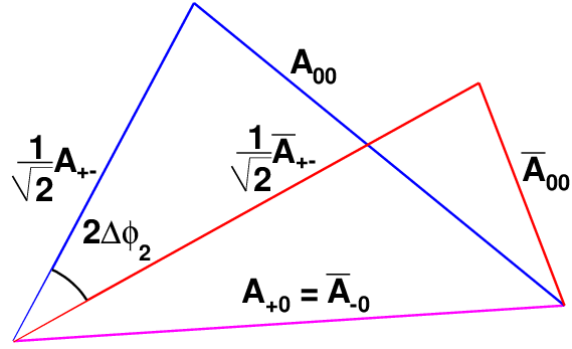}
  \caption{A sketch of the isospin triangle.}
  \label{fig_iso}
\end{figure}

The CKM angle $\phi_2$ can be determined by measuring the time-dependent asymmetry between \Bz\ and \Bzb\ decays into a common $CP$ eigenstate~\cite{CP} made out of unflavored quarks ($\bar{b} \rightarrow \bar{u}u\bar{d}$ quark transitions). Examples are the decays $\Bz \rightarrow \pi\pi$, $\rho\pi$, $\rho\rho$ and $\aone\pi$~\cite{pipi_Belle,pipi_BaBar,pipi_LHCb,rhoprhom_BaBar, rhoprhom_Belle1, rhoprhom_Belle2,belle_r0r0,r0r0Babar}. In the decay sequence, $\YFS \rightarrow B_{CP}B_{\rm Tag} \rightarrow f_{CP}f_{\rm Tag}$, where one of the $B$ mesons decays into a $CP$ eigenstate  $f_{CP}$ at a time $t_{CP}$ and the other decays into a flavor specific final state $f_{\rm Tag}$ at a time $t_{\rm Tag}$, the time-dependent decay rate is given by
\begin{equation}
  {P}(\Delta t, q) = \frac{e^{-|\Dt|/\tau_{B^0}}}{4\tau_{B^0}} \bigg[ 1+q(\Acp\cos\Delta m_d \Dt + \Scp\sin\Delta m_d \Dt) \bigg],
\label{eq1}
\end{equation}
where $\Dt \equiv t_{CP}- t_{\rm Tag}$ is the lifetime difference between the two $B$ mesons, $\Delta m_d$ is the mass difference between the mass eigenstates  $B_{H}$ and $B_{L}$ and $q = +1 (-1)$ for $B_{\rm Tag} = \Bz (\Bzb)$. The $CP$ asymmetry is given by 
\begin{equation}
\frac{N(\bar{B}\to f_{CP}) - N(B\to f_{CP})}{N(\bar{B}\to f_{CP}) + N(B\to f_{CP})},
\label{eq_asym}
\end{equation}
where $ N(B(\bar{B})\to f_{CP})$ is the number of events of a $B(\bar{B})$ decaying to $f_{CP}$, the asymmetry can be time-dependent.
 The parameters \Acp\ and \Scp\ describe direct and mixing-induced $CP$ violation, respectively~\footnote{There exists an alternate notation where $C_{CP} = -\Acp$.}. \\
 At tree level one expects $\Acp=0$ and $\Scp=\sin2\phitwo$ for the above mentioned decays sensitive to $\phi_2$. Possible penguin contributions can give rise of direct $CP$ violation, $\Acp\neq 0$ and also pollute the measurement of \phitwo, $\Scp=\sqrt{1-\Acp^{2}}\sin(2\phitwo^{eff})$ where the observed $\phitwo^{eff} \equiv \phitwo - \Delta \phitwo$ is shifted by $\Delta \phi_2$ due to different weak and strong phases from additional non-leading contributions.
\begin{figure}[htb]
  \centering
  \includegraphics[width=.45\textwidth]{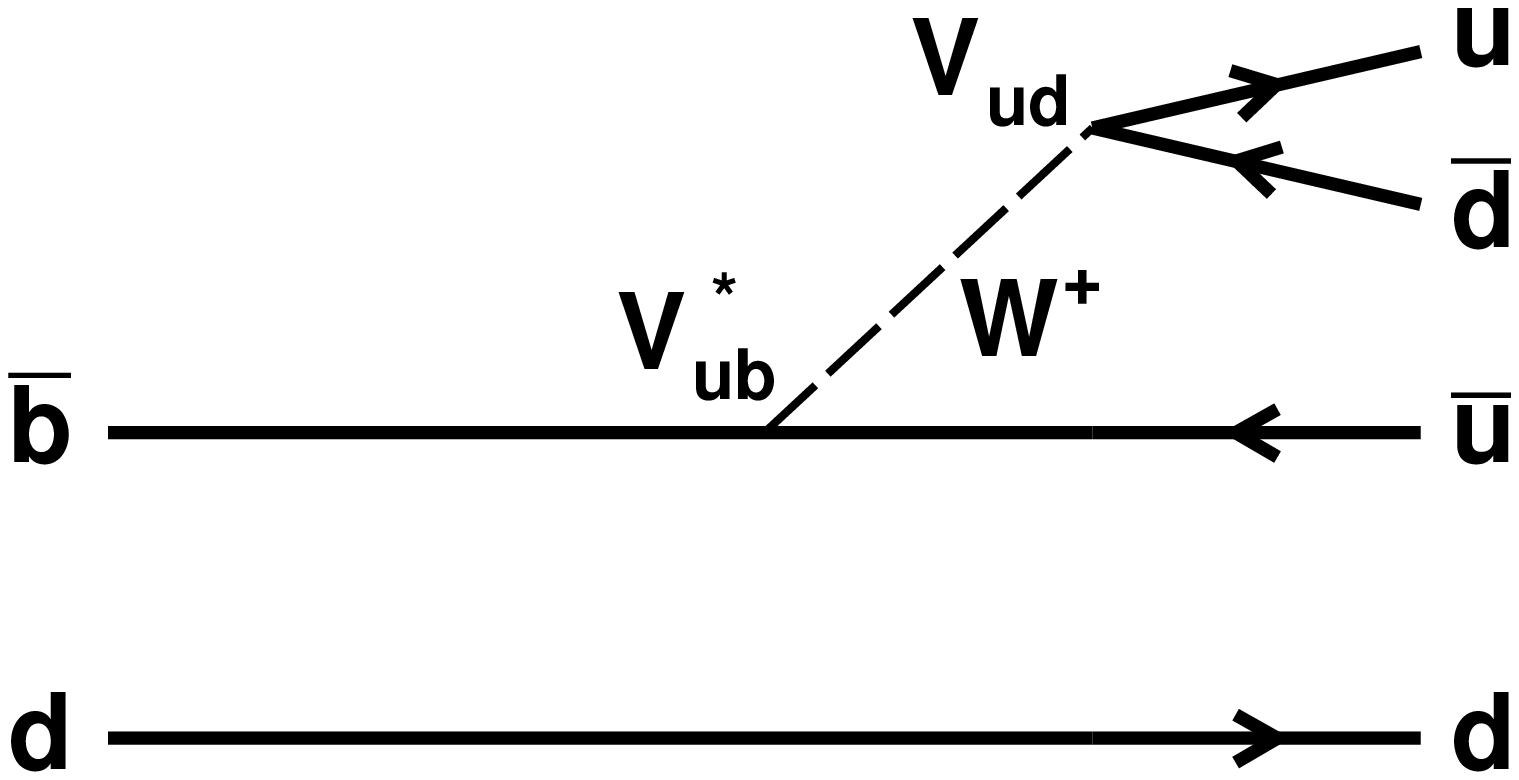}
  \includegraphics[width=.45\textwidth]{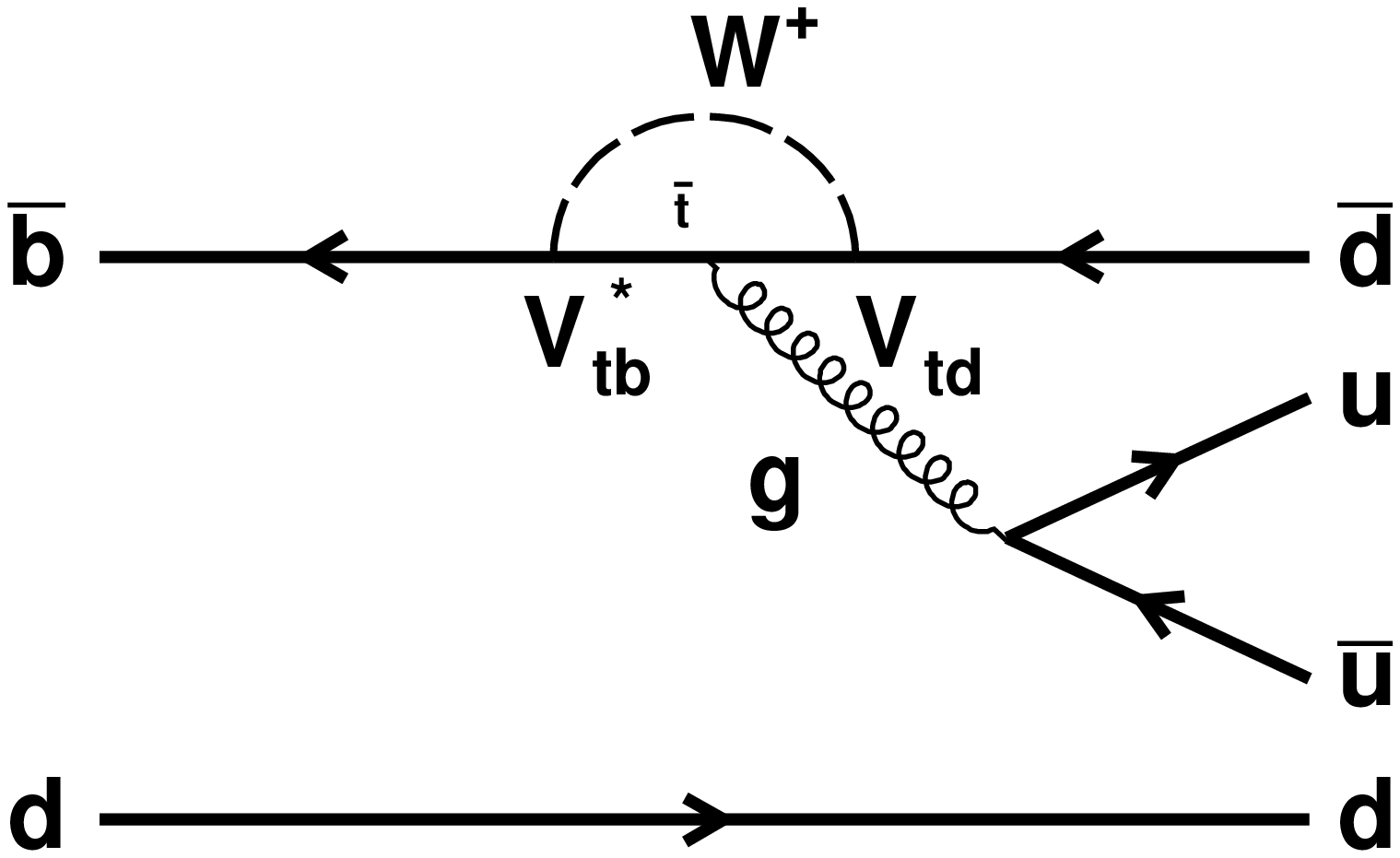}
 \put(-380,1){\scriptsize a)}
 \put(-180,1){\scriptsize b)}\\
  \caption{a) leading order and b) penguin feynman diagrams for color-allowed $b \rightarrow u \bar{u} d$ transitions.}
  \label{fig_pipi_feyn}
\end{figure}
Despite this, it is possible to determine $\Delta \phitwo$ in $\Bz \rightarrow h^{+} h^{-}$ with an $SU(2)$ isospin analysis by considering the set of three $B \rightarrow hh$ decays where the $hh$s are either two pions or two longitudinally polarized $\rho s$, related via isospin symmetry~\cite{iso}. The $B \rightarrow h^{i} h^{j}$ amplitudes $A_{ij}$ obey the triangle relations,
\begin{equation}
  A_{+0} = \frac{1}{\sqrt{2}}A_{+-} + A_{00}, \;\;\;\; \bar{A}_{-0} = \frac{1}{\sqrt{2}}\bar{A}_{+-} + \bar{A}_{00}.
  \label{eq_iso}
\end{equation}
Isospin arguments demonstrate that $\Bp \rightarrow h^{+} h^{0}$ is a pure first-order mode in the limit of neglecting electroweak penguins, thus these triangles share the same base, $A_{+0}=\bar{A}_{-0}$, see Fig.~\ref{fig_iso} for an illustration. $\Delta \phitwo$ can then be determined from the difference between the two triangles. This method has an inherent four-fold discrete ambiguity in the determination of $\sin(2\phitwo)$.
%All results presented in this section are preliminary.

\subsection{The Decay $B^0\to \rho^+\rho^-$}
\begin{figure}[htb]
  \centering
 \includegraphics[height=!,width=0.6\columnwidth]{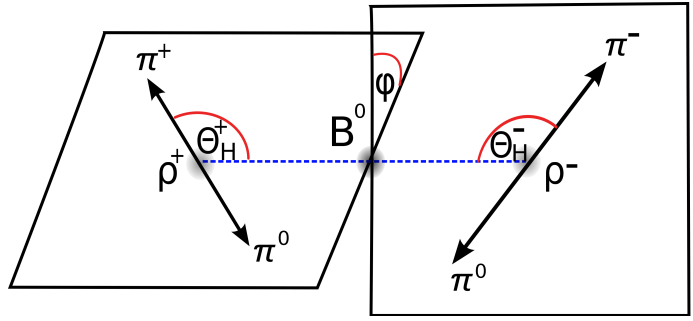}
 \caption{The helicity angles $\cos\Theta_{\rm H}^{\pm}$; each one is defined in its $\rho$ rest frame.}
  \label{fig_hel}
\end{figure}

\begin{figure}[htb]
  \centering
\includegraphics[height=!,width=0.32\columnwidth]{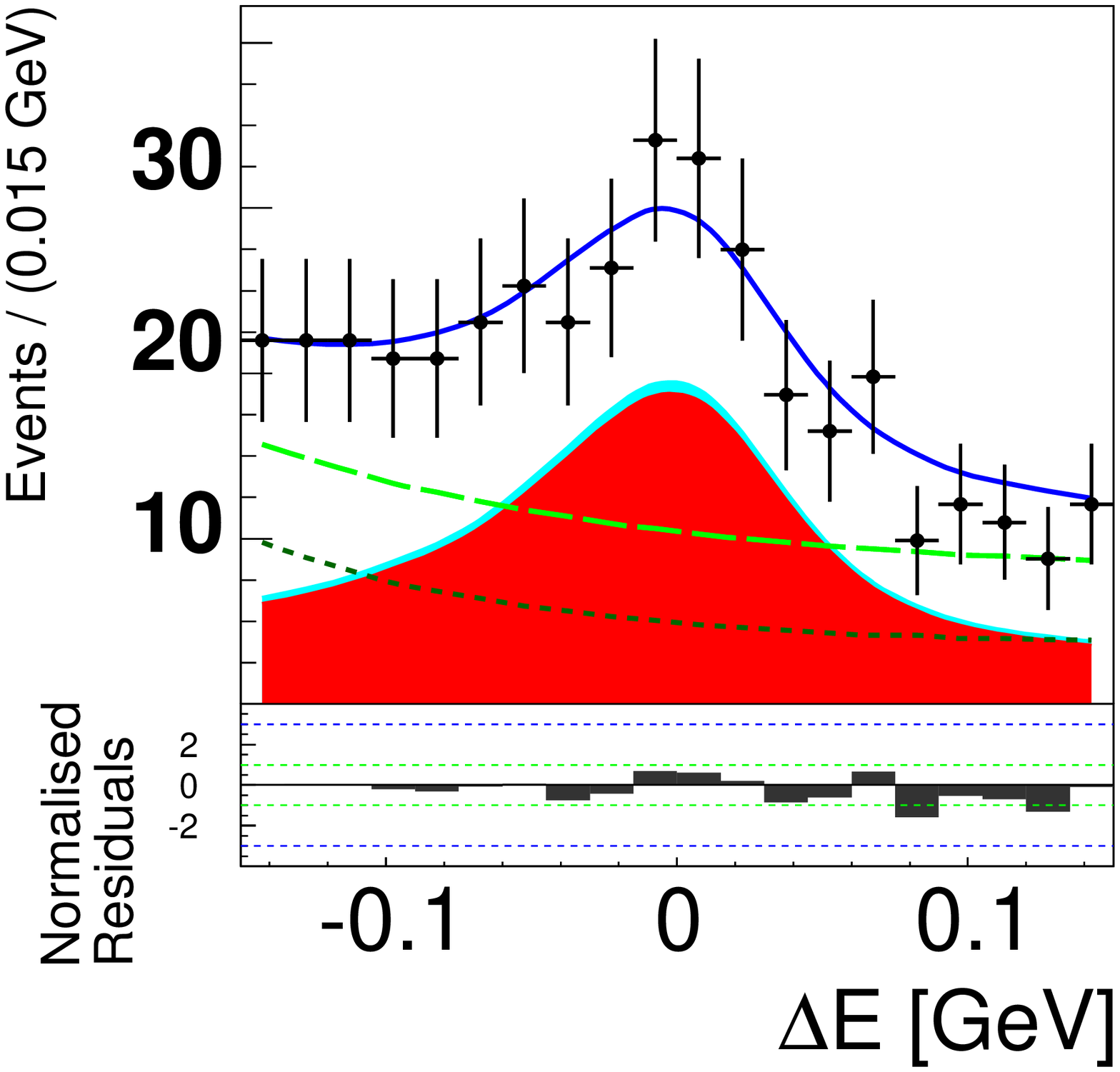}                                                              
\includegraphics[height=!,width=0.32\columnwidth]{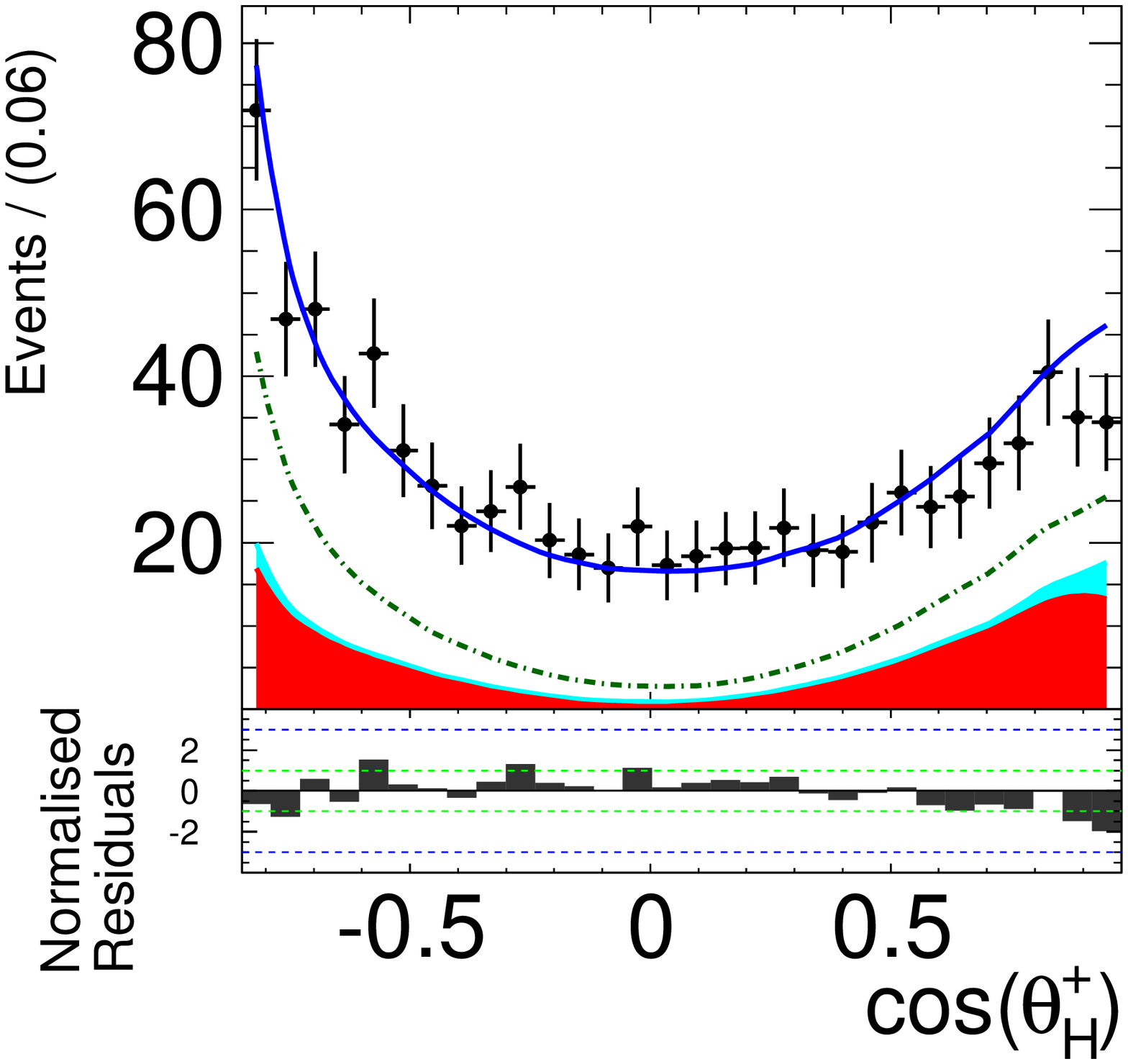}                                                             
\includegraphics[height=!,width=0.32\columnwidth]{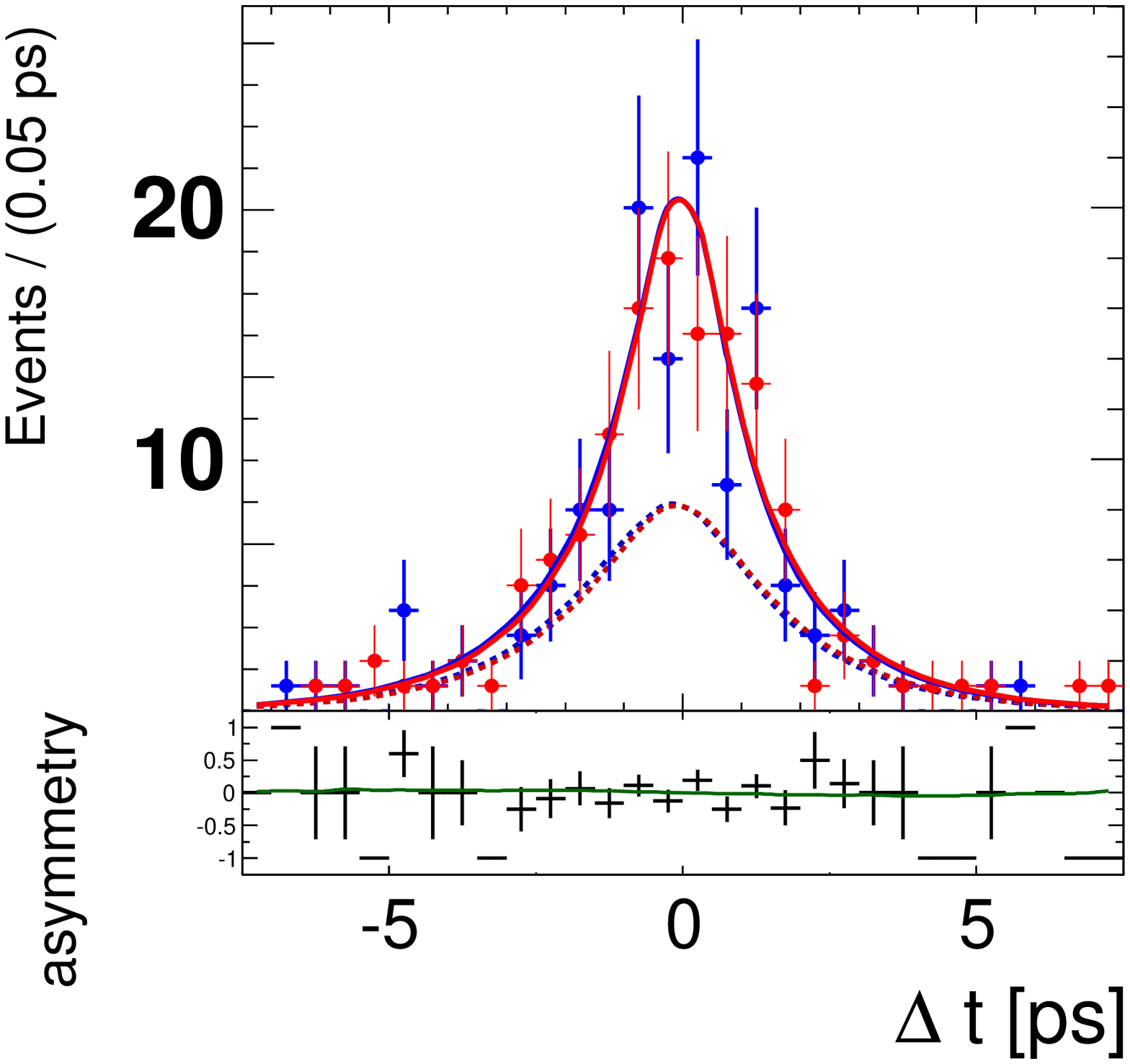}                                                         
\put(-50,105){\textcolor{blue}{\footnotesize $q=+1$}}\put(-50,95){\textcolor{red}{\footnotesize $q=-1$}} 
\put(-195,110){\footnotesize  \textcolor{red}{signal}} \put(-195,102){\footnotesize  \textcolor{cyan}{$B\to 4\pi$}}\put(-195,94){\footnotesize  \textcolor{mygreen}{$B\bar{B}$}}\\
  \caption{ Signal enhanced distributions of (a) $\Delta E$, (b) $\cos\Theta_H$, and (c) $\Delta t$ for the two flavors of $B_{\rm Tag}$ ($B_{\rm Tag} = B^0$ for $q=+1$) with the fit result on top. The shaded red area is the $B^0\to\rho^0\rho^0$ contribution. Furthermore, all $B$ decays with a four pion final state are shown in cyan, the entire ($B\bar{B}$) background in dashed (dash-dotted dark) green and the full PDF in blue.}
  \label{p_r0r0}
\end{figure}

\begin{figure}[htb]
  \centering
 \includegraphics[height=!,width=0.65\columnwidth]{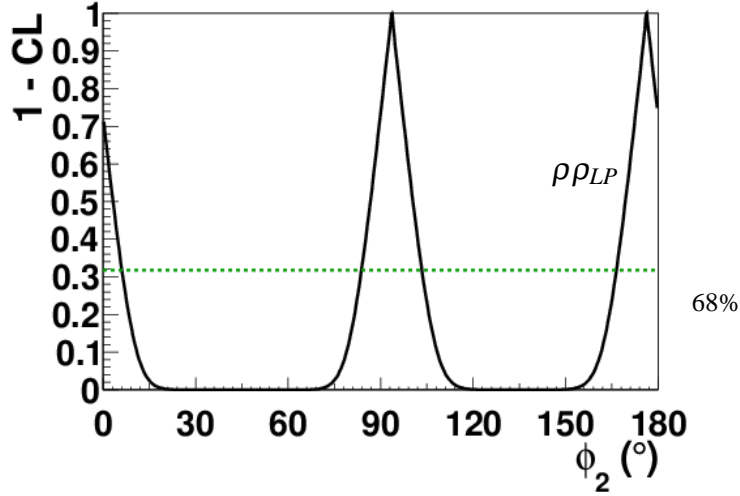}
 \put(-15, 70){\small \textcolor{black}{ $68\%$}}                                                                                                           
    \put(-68, 120){\small \textcolor{black}{\large $\rho\rho_{LP}$}}\\          
 \caption{Probability scan of $\phi_2$ in the $B\to\rho\rho$ system.}
  \label{fig_phi2}
\end{figure}

 Having a decay into two vector particles, an angular analysis is performed to separate the $CP$-even states from the  $CP$-odd states for the isopsin analysis. Longitudinal polarized states correspond to pure $CP$-even states and their fraction, $f_L$, is obtained from a fit to the cosine of helicity angles, $\Theta_{\rm H}^{\pm}$, which are defined as sketched in Fig.~\ref{fig_hel}. Previous measurements show that $f_L$ is consistent with one~\cite{rhoprhom_BaBar, rhoprhom_Belle1, rhoprhom_Belle2}, consequently the isospin analysis is performed for longitudinal polarization only.
In addition to combinatorial background, the presence of multiple, largely unknown backgrounds with the same four-pion final state as $B^0\to \rho^+\rho^-$ make this decay quite difficult to isolate and interference between the various $4\pi$ modes needs to be considered. Besides updating to the full data set, the branching fraction, the fraction of longitudinally polarized $\rho$ mesons on this decay and the $CP$ violating parameters are obtained simultaneously from the fit to the data. In the fit, Belle uses the missing energy $(\Delta E \equiv E_{B}^{*}-E_{\rm beam}^*)$, the beam-constraint $B$ mass ($M_{\rm bc}=\sqrt{E_{\rm beam}^{*2} - p_{B}^{*2}})$, the masses and helicity angles of the two reconstructed $\rho^\pm$ mesons, a fisher discriminant to separate the jet-like $e^+e^-\to q\bar{q},\; (q=u,\;d,\;s,\;c)$ background from the spherical $B\bar{B}$ decays and the $\Delta t$ distribution for the two flavors of $B_{\rm Tag}$. 
They obtain
\begin{itemize}
\item[]${\cal B}(B^0\to\rho^+\rho^-)=(28.3\pm 1.5\;(\rm stat) \pm 1.4\;(\rm syst))\times 10^{-6},$
\item[]$f_L = 0.988\pm\; 0.012\;(\rm stat ) \pm\;0.023(\rm syst),$
\item[]${S}_{CP} =-0.13\pm0.15(\rm stat ) \pm0.05\;(\rm syst)$ and 
\item[]${A}_{CP} =0.00\pm0.10(\rm stat ) \pm0.06\;(\rm syst)$.
\end{itemize}
This is currently the most precise measurement of the branching fraction and polarization of $B\to \rho^+\rho^-$ decays as well as the tightest constraint on $CP$ violation in this mode.
 Fig.~\ref{p_r0r0} shows the projections onto $\Delta E$, $\cos\Theta_{\rm H}^{+}$ and onto $\Delta t$ for the two flavor of $B_{\rm Tag}$, each with the fit result on top.
The results from this measurement are used in an isospin analysis together with other Belle measurements~\cite{belle_r0r0, rpr0_Belle} (longitudinal polarization only).  Fig.~\ref{fig_phi2}~b) shows the \phitwo\ scan from the isospin analysis, the constraint most consistent with other measurements of the CKM triangle is $\phitwo = (93.7 \pm 10.6)^{\circ}$ and the penguin pollution is consistent with zero: $\Delta\phi_2 = (0.0\pm9.6)^{\circ}$. In the $B\to\rho\rho$ system, the relatively small amplitude of $B^0\to\rho^0\rho^0$ makes the isospin triangles flat and therefore the isospin analysis has no ambiguity. %compared for example to the current status of $\phi_2$ from $B\to \pi\pi$.

\subsection{The Decay $B^0\to\piz\piz$}
\begin{figure}[htb]
\centering
\includegraphics[height=!,width=0.9\columnwidth]{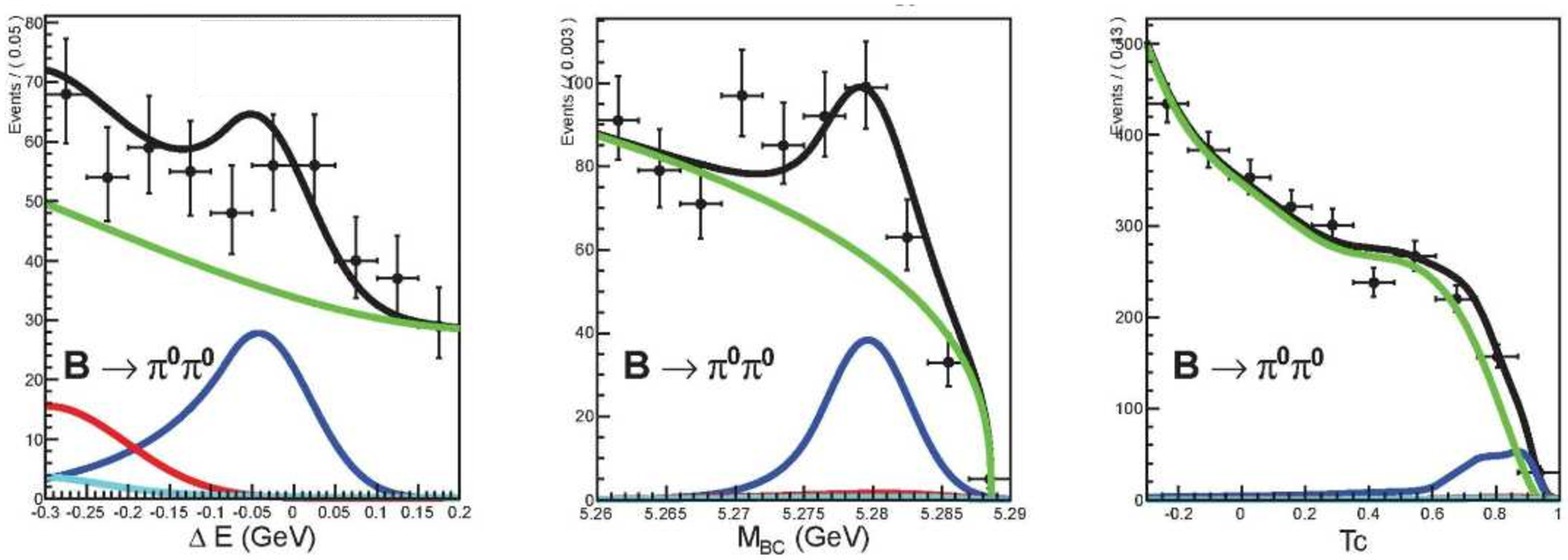}
%\put(-261,90){\mycbox{white}}
 \caption{ Signal enhanced distributions of (a) $\Delta E$, (b) $\cos\Theta_H$ and (c) $\Delta t$ with the fit result on top. Contributions from signal, continuum, $\rho\pi$ and other rare B decays are shown by blue, green, red and cyan respectively.}
  \label{p_p0p0}
\end{figure}

This decay is an important input for the isospin analysis in the $B\to\pi\pi$ system. Being reconstructed from four $\gamma$s makes this measurement experimentally quite challenging. A fit to \De, \Mbc\ and a fisher discriminant $T_C$ is performed and a preliminary branching fraction of 
\begin{itemize}
\item[] ${\cal B}(B\to\piz\piz) = (0.9\pm 0.12\;{\rm stat}\pm 0.10\;{\rm syst})\times 10^{-6}$
\end{itemize}
 is obtained. Signal enhanced projections are shown in Fig.~\ref{p_p0p0}. It is planned to supersede this currently most precise measurement of the branching fraction with one including the determination of the direct $CP$ violation in this mode.\\
The upcoming Belle 2 experiment will make a time-dependent analysis possible, as the accumulated data will provides enough data to use converted photons to determine the $B$ vertex. Including the mixing-induced $CP$-violation parameter of $B\to\piz\piz$ in the isospin analysis might remove the four-fold ambiguity from the isospin analysis being currently present in the $B\to\pi\pi$ system.

\section{The Decay \BtoDK\ and the CKM Angle $\phi_3$}

\begin{figure}[htb]
\centering
\includegraphics[height=!,width=0.99\columnwidth, bb=2 500 1300 830]{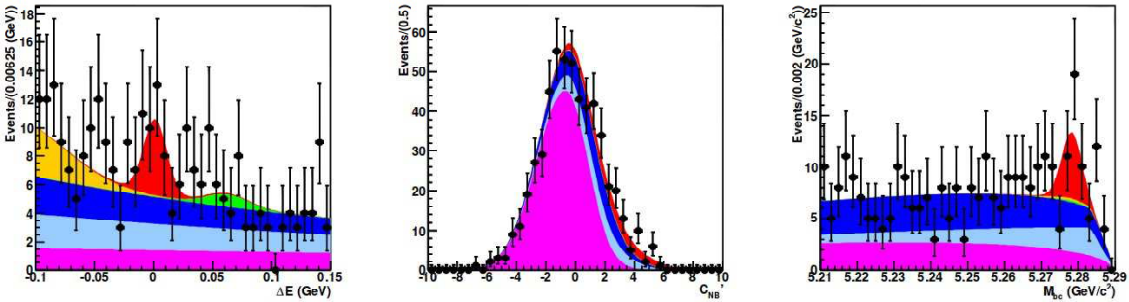}
 \caption{ Signal enhanced distributions of (a) $\Delta E$, (b) $\cos\Theta_H$ and (c) $\Delta t$ with the fit result on top. The signal contribution is shown in red.}
  \label{p_DK}
\end{figure}

\begin{figure}[htb]
\centering
\includegraphics[height=!,width=0.7\columnwidth, bb = 20 330 1150 900]{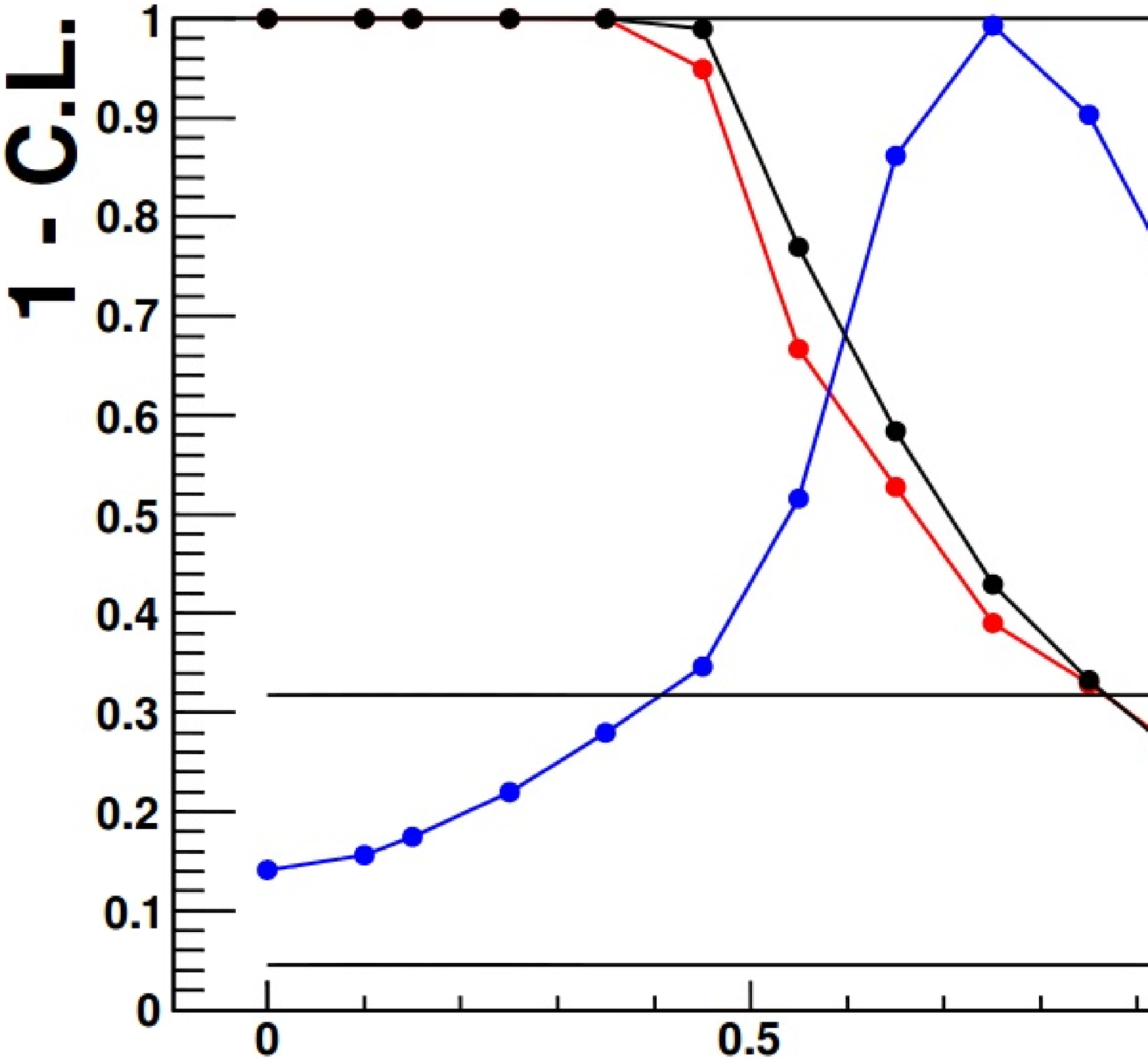}
 \caption{Probability scan of $r_S$.}
  \label{p_DK2}
\end{figure}

A model independent dalitz plot analysis~\cite{phi3_theo, B2DK_Belle} of the decay \BtoDK\ has been performed for the first time. The flavor-specific decay, $K^{*0}\to K^+\pim$, allows to determine the flavor of the $B$ meson. The number of events in different bins of the dalitz plot of the $D$ meson coming from either a $B^0$ ($N_i^+$) or a $\bar{B}^0$ ($N_i^-$) is given by
\begin{equation}
N_i^\pm = h_B[K_{\pm i} + r_S^2K_{\mp i} + 2k\sqrt{K_iK_{-i}}(x_\pm c_i \pm y_\pm s_i))],
\label{e_BDK}
\end{equation}
where $h_B$ is a normalization constant, $K_{\pm i}$ are the entries in the $D$ (+) or $\bar{D}$ (-) dalitz plot, $k=0.95 \pm 0.03$~\cite{B2DK_Babar} accounts for interference effects in the $D$ decay, and $c_i$ and $s_i$ include information on the average of the phase variation within a dalitz plot bin. All information on the $D$ dalitz plot is provided by measurements from the CLEO collaboration~\cite{Cleo_D}. The observables $x_\pm\equiv r_{s}\cos(\delta_S \pm \phi_3)$ and $y_\pm \equiv r_{s}\sin(\delta_S \pm \phi_3)$ from the interference term in Equ.~\ref{e_BDK} allow an extraction of the CKM angle $\phi_3$ in general, where the ratio between the cabbibo-allowed and double-cabbibo-suppressed amplitudes, $r_S\equiv\frac{\bar{A}}{A}=\frac{A(\bar{B}\to \bar{D}^0\bar{K}^{*0})}{A(\bar{B}\to D^0\bar{K}^{*0})}$, indicates the sensitivity to $\phi_3$. Belle obtains
\begin{eqnarray}
x_+=+0.1^{+0.7+0.0}_{-0.4-0.1}\pm0.1, &\;\;\;\;\;\;\;y_+=+0.3^{+0.5+0.0}_{-0.8-0.1}\pm0.1,\nonumber \\
x_-=+0.4^{+0.5+0.0}_{-0.8-0.1}\pm0.0, &\;\;\;\;\;\;\;y_-=-0.6^{+0.8+0.1}_{-1.0-0.0}\pm0.1,\nonumber
\end{eqnarray}
and uses the result to obtain an upper limit of 
\begin{eqnarray}
r_S<0.87 \nonumber 
\end{eqnarray}
at the one $\sigma$ level. The (dalitz plot integrated) fit results are shown in Fig.~\ref{p_DK} and the $r_S$ scan is shown in Fig.~\ref{p_DK2}. This mode is still statistically limited but will give additional insights on $\phi_3$ when the Belle 2 data will be available.

\section{Summary}
We have presented three recent and preliminary measurements from Belle sensitive to the CKM phases \phitwo\ and $\phi_3$ using the full data set of $772$ million $B\bar{B}$ pairs .
Measurements of the branching fraction, the polarization and the $CP$ asymmetries in $B \rightarrow \rho^+\rho^-$ were used to update the $\phi_2$ isospin constraint from Belle. The branching fraction measurement of of $B\to\pi^0\pi^0$ has been presented and the importance for this mode the isospin analysis in the $B\to \pi\pi$ system has been discussed. The current world averages of \phitwo\ as computed by the CKMfitter~\cite{CKMfitter} and UTfit~\cite{UTfit} collaborations are $\phitwo = (87.6^{+3.5}_{-3.3})^{\circ}$ and $\phitwo = (88/6 \pm 3.3 )^{\circ}$, respectively. Furthermore we presented a preliminary measurement of \BtoDK\ decays and discussed the sensitivity to $\phi_3$. All shown results are in good agreement with other SM based constraints on the CKM triangle. With Belle 2 being built~\cite{Belle2} and the LHCb operating, the next generation of $B$ physics experiments are expected to further reduce the uncertainty of the CKM observables and might reveal new phenomena.

\end{document}